\documentclass[%
unsortedaddress,
preprint,
 amsmath,amssymb,
 aps,
pre
]{revtex4-1}

\usepackage{graphicx}
\usepackage{dcolumn}
\usepackage{bm}
\usepackage{hyperref}

\begin{document}

\title{Motile dissenters disrupt the flocking of active granular matter}

\author{Pradip K. Bera and A. K. Sood}
\email{asood@iisc.ac.in}

\affiliation{Department of Physics, Indian Institute of Science, Bangalore 560012, India}

\date{\today}

\begin{abstract}
We report flocking in the dry active granular matter of millimeter-sized two-step-tapered rods without an intervening medium. The system undergoes the flocking phase transition at a threshold area fraction $\sim 0.12$ having high orientational correlations between the particles. However, the one-step-tapered rods do not flock and are used as the motile dissenters in the flock-forming granular matter. At the critical fraction of dissenters $\sim 0.3$, the flocking order of the system gets completely destroyed. The variance of the system's order parameter shows a maximum near the dissenter fraction $f \sim 0.05$, suggesting a finite-size crossover between the ordered and disordered phases.
\end{abstract}

\maketitle

\section{INTRODUCTION}
The effect of static defects or motile non-aligning agents (called dissenters \cite{yllanes2017many}) on the flocking or other collective motions of a group is a subject of recent interest in model systems and simulations \cite{chepizhko2013optimal,berdahl2013emergent,quint2015topologically,pincce2016disorder,morin2017distortion,sandor2017dynamic,reichhardt2017collective,reichhardt2018clogging,das2018polar,ai2019flow,guttal2010social,couzin2011uninformed,mccandlish2012spontaneous,baglietto2013gregarious,ariel2015order}. For example, static disorder in the form of physical obstacles is found to alter the group's motion dramatically by bringing spatial heterogeneity in the system. On the other side, motile dissenters lacking the aligning property can disrupt the global consensus and destabilize the flocking state. Recent numerical simulations by Copenhagen et al. \cite{copenhagen2016self} and Yllanes et al. \cite{yllanes2017many} have shown that the flocking state of aligners can be disrupted completely at a certain fraction of dissenters. To date, there are no experiments to study the role of dissenters in flocking behavior, a motivation for our present study.

\begin{figure}
\includegraphics[width=0.5\textwidth]{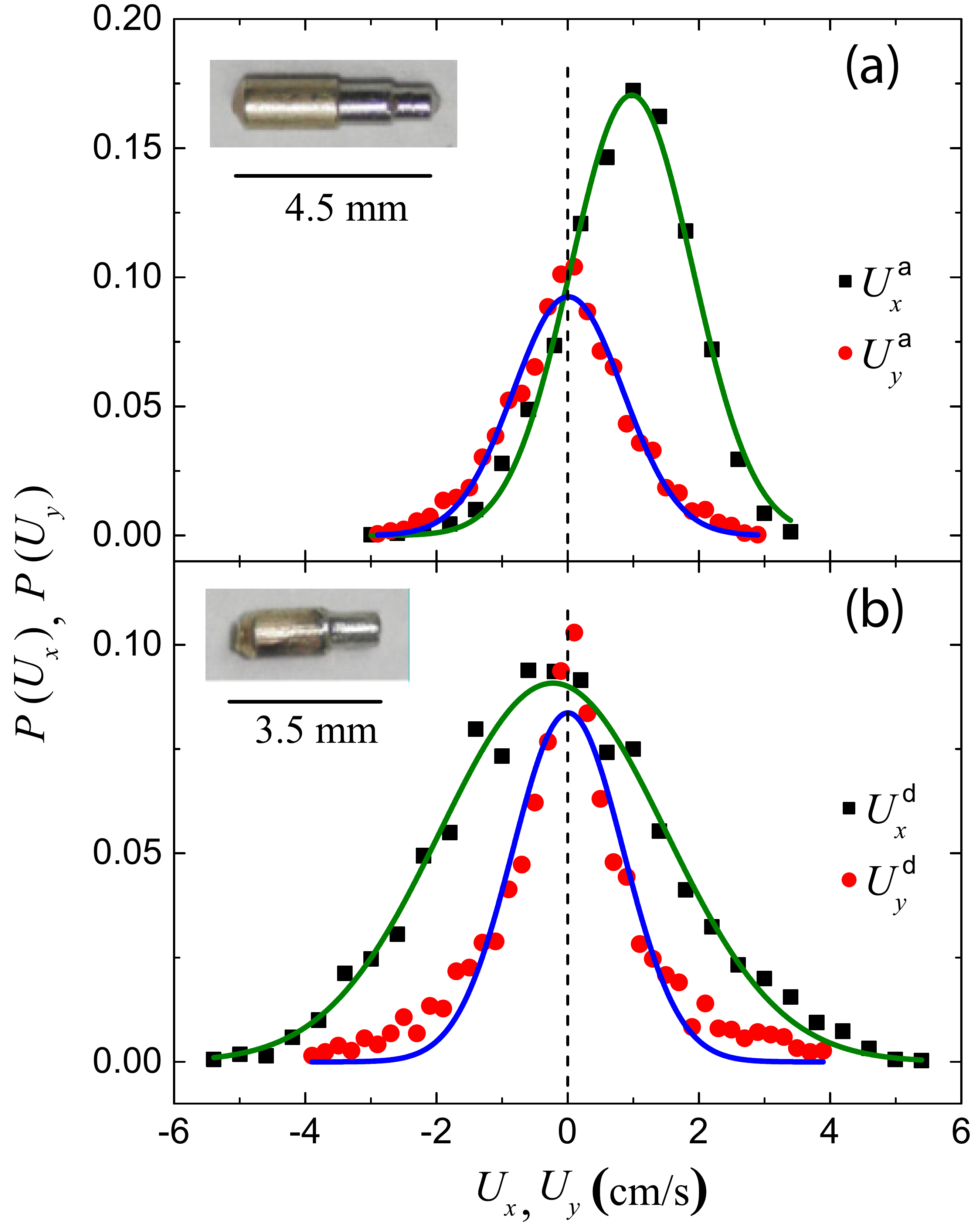}
\caption{Distribution of velocity of a single particle, along the polarity direction ($U_x$) and orthogonal to that ($U_y$). (a) Plots for two-step-tapered polar active rod (called aligner), (b) for one-step-tapered polar active rod (called dissenter) and the corresponding rod images are shown. Solid curves are gauss fits with peak positions $\sim 1.0, \:0.0$ cm/s in (a) and $\sim -0.2$ cm/s ($\sim$ resolution limit), $\sim 0.0$ cm/s in (b). Dotted lines represent zero velocity. Shaking strength $\Gamma = 4$ and frequency $n = 200$ Hz; are fixed throughout our study.}
\label{F1}
\end{figure}

We have used two types of brass particles having different degree of polarity (Fig.~\ref{F1}). Our experimental observations are as follows: (i) Flocking in a granular medium of two-step-tapered polar active agents (called aligners) is observed above a certain area fraction without any intervening medium. (ii) One-step-tapered polar active agents (called dissenters) do not flock even at a very high area fraction. (iii) The mixed systems of flock-forming aligners and dissenter particles show flocking with a low fraction of dissenters. However, above a critical dissenters' fraction, the mixed system does not flock, similar to the simulation results \cite{copenhagen2016self}. We quantify orientational correlations and other measures as the fraction of dissenters is increased.

\section{EXPERIMENTAL DETAILS}
Our active granular material is a collection of macroscopic polar brass rods vibrated vertically by a magnetic shaker (LDS V406-PA100E). The aligner-rods are 4.5 mm long (denoted by $L_\mathrm{a}$), with diameter tapered in two steps from 1.1 mm at the thick end to 0.7 mm at the thin end. On the other hand, the dissenter-rods are 3.5 mm long, with diameter tapered in one step from 1.1 mm at the thick end to 0.7 mm at the thin end. Our monolayer of rods is confined in the 1.2 mm gap between the flower-shaped experimental cell and its top glass lid (see Supplemental Material \cite{supmat} for details about the setup calibration). We keep the shaker oscillation amplitude ($\mathcal{A}= 0.025$ mm) and frequency ($n=200$ Hz) fixed during our experiments. The non-dimensional shaking strength ($(2 \pi n)^{2} \mathcal{A} /{g}$; $g$ is gravity) is 4.0 (also measured by the two orthogonally attached MPU-6050 accelerometers). Both types of rods imitate self-propulsion by transducing the vertical vibrations into fluctuating but persistent horizontal motion, in the tail-to-head direction \cite{yamada2003coherent,narayan2007long,kumar2011symmetry,kumar2019trapping}. A Redlake MotionPro X3 camera is used to capture images at 30 fps (frames per second) during studies with a single particle and at 1 fps during studies with the collection of particles. Fiji (ImageJ) is used for image analysis \cite{schindelin2012fiji} (see Supplemental Material \cite{supmat}).

\section{RESULTS}
We disperse the required number of particles in the clean cell and keep the shaker on for 500 sec to observe the onset of the flocking transition over time. We repeat this procedure three times with a given set of particles to get good statistical estimations. We will first present the collective behavior of only aligners before presenting the effect of dissenters on the flocking transition. 

Figure~\ref{F1} shows the single-particle dynamics of aligners and dissenters, present alone in the cell (Movies S2 and Movies S3 respectively). Velocity components $U_x$ and $U_y$ are calculated by taking the projection of the laboratory frame velocity to the particle polarity direction ($U_x$) and taking projection orthogonal to the polarity direction ($U_y$). The statistical anisotropy of the dynamics is evident from the probability distributions of $U_x$ and $U_x$ for both the particles, which shows a much greater dispersion along the rod axis direction than transverse to it. For the aligner, the $U_x$ component shows a maximum in the probability distribution, $P(U_x)$ at $\sim 1$ cm/s whereas $P(U_y)$ has a peak close to zero. For the dissenter, $P(U_x)$ shows a peak at $\sim -0.2$ cm/s, close to the resolution limit whereas $P(U_y)$ is peaked close to zero. Both velocity components of aligners and dissenters show Gaussian behavior. The aligners always trace out a finite displacement after some time in the forward direction ($+U_x$ direction) but the dissenter's movement is highly non-directional.

\begin{figure}
\includegraphics[width=0.48\textwidth]{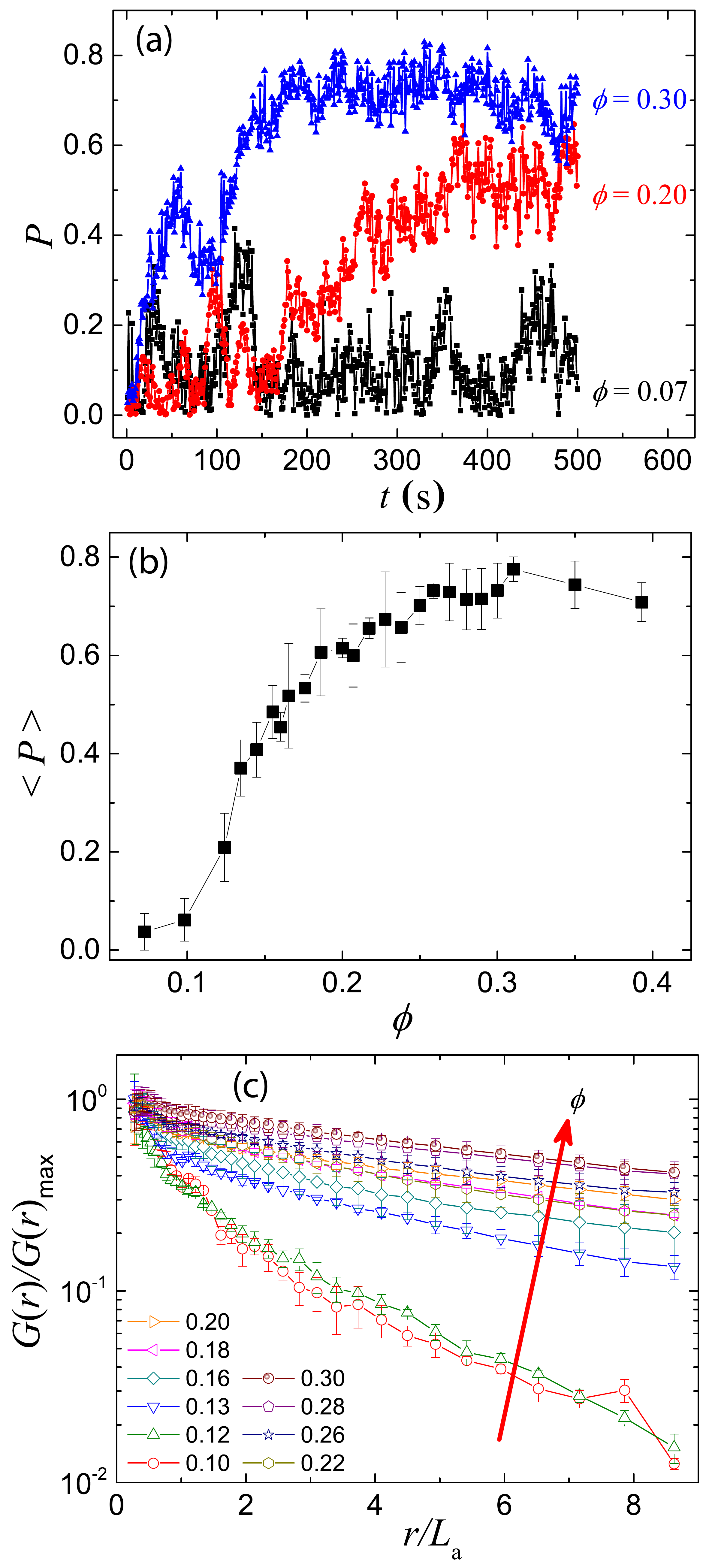}
\caption{The system with only aligners; error bar with each data point represents the standard deviation of the quantity in three repeated experiments. (a) Flocking order parameter ($P(t)$) vs time ($t$) is shown for three different area fractions ($\phi$). (b) Average steady-state value of $P(t)$ ($\left\langle P \right\rangle$; averaged considering steady-state frames and then over the three repeated experiments) is plotted against $\phi$. (c) Average orientational correlation function ($G(r)/G(r)_\mathrm{max}$) vs inter-particle separation ($r/L_\mathrm{a}$; $L_\mathrm{a}$ is aligner's length) is plotted for different $\phi$. For the given $\phi$ and $r$, $G(r)$ is averaged over satisfying pairs in the steady-state frames and then over the three repeated experiments. The red arrow is towards increasing $\phi$.}
\label{F2}
\end{figure}

The collective behavior of the system with only aligners is markedly different from that of the system with only dissenters (see Supplemental Material \cite{supmat}, Fig.~S3, Movies S4, S5, S6, S7). At low area fraction $\phi=0.07$ ($\phi$ = area covered by the two-dimensional projections of the rods/flower area), aligners show a completely isotropic disordered state. Above a threshold area fraction ($\phi=0.12$), they form a dynamically steady flock where most of the rods are aligned along the flocking direction. This will be discussed later. Systems with only dissenters do not show flocking even at a high area fraction up to $\phi=0.70$. For $\phi > 0.70$, the system shows an active jammed state \cite{henkes2011active}. We quantify (following Ref. \cite{kumar2014flocking}) the flocking order for each time frame by resolving in-plane rod's orientation, $\textbf{n}_i$, into local polar-coordinate components with the coordinate center located at the flower center and define ($\textbf{n}_i \cdot \textbf{r}_i$, $\textbf{n}_i \times \textbf{r}_i$) $\equiv$ $\textbf{P}_i$, where $\textbf{r}_i$ is the unit radial position vector of the $i$th particle. For each image frame, we then calculate $P \equiv \left|\left\langle \textbf{P}_i \right\rangle\right|$, averaged over all particles. Figure~\ref{F2}(a) shows the growth kinetics of the measured flocking order parameter ($P(t)$) for the system having only aligners for three area fractions. For $\phi=0.07$, the system shows a completely isotropic disordered state with $P(t)$ fluctuating near zero, implying disordered state. For $\phi=0.20$ and 0.30, aligners form a dynamically stable flock within 300 sec. The order parameter grows from a random configuration and reaches a steady saturation value close to 1. The average steady-state order parameter ($\left\langle P \right\rangle$) is calculated by taking an average of $P$ over the steady-state time frames and then over the three repeated experiments for each $\phi$ (Fig.~\ref{F2}(b)). The error bar is the standard deviation in $\left\langle P \right\rangle$ over three repeated experiments for the same $\phi$. The same calculation strategy is adopted for all $\left\langle P \right\rangle$ presented in the subsequent plots. For all the area fractions below 0.13, aligners do not form a dynamically stable flock. Aligners form a dynamically stable flock for the area fractions $0.13 \leq \phi \leq 0.40$. The smooth variation in order parameter above $\phi = 0.12$ is due to the finite system size effects as the discontinuous flocking transition is observed in the finite but larger system sizes in simulations \cite{chate2019dry}. For $0.40 < \phi \leq 0.70$, the randomly running active matter condensates to a large single cluster of particles and eventually goes to a dynamically jammed state \cite{reichhardt2014absorbing}.
For $\phi >$ 0.70, we observe the active jammed state. The condensation and the jamming are not pursued in this work. We now explore how the orientational correlation function grows as $\phi$ increases. In Fig.~\ref{F2}(c) we plot the average steady-state orientational correlation function $G(r)/G(r)_\mathrm{max} = \left\langle\textbf{P}_i(0) \cdot \textbf{P}_j(r)\right\rangle_\mathrm{all \:pairs}$, a measure of the probability that two rods separated by distance $r$ are pointing in the same direction, and the data is averaged over various pairs in the steady-state frames and in the three repeated experiments. We observe that above $\phi=0.12$, the system shows long-range correlation (also see Supplemental Material \cite{supmat}, Fig.~S4).

\begin{figure*}
\includegraphics[width=0.90\textwidth]{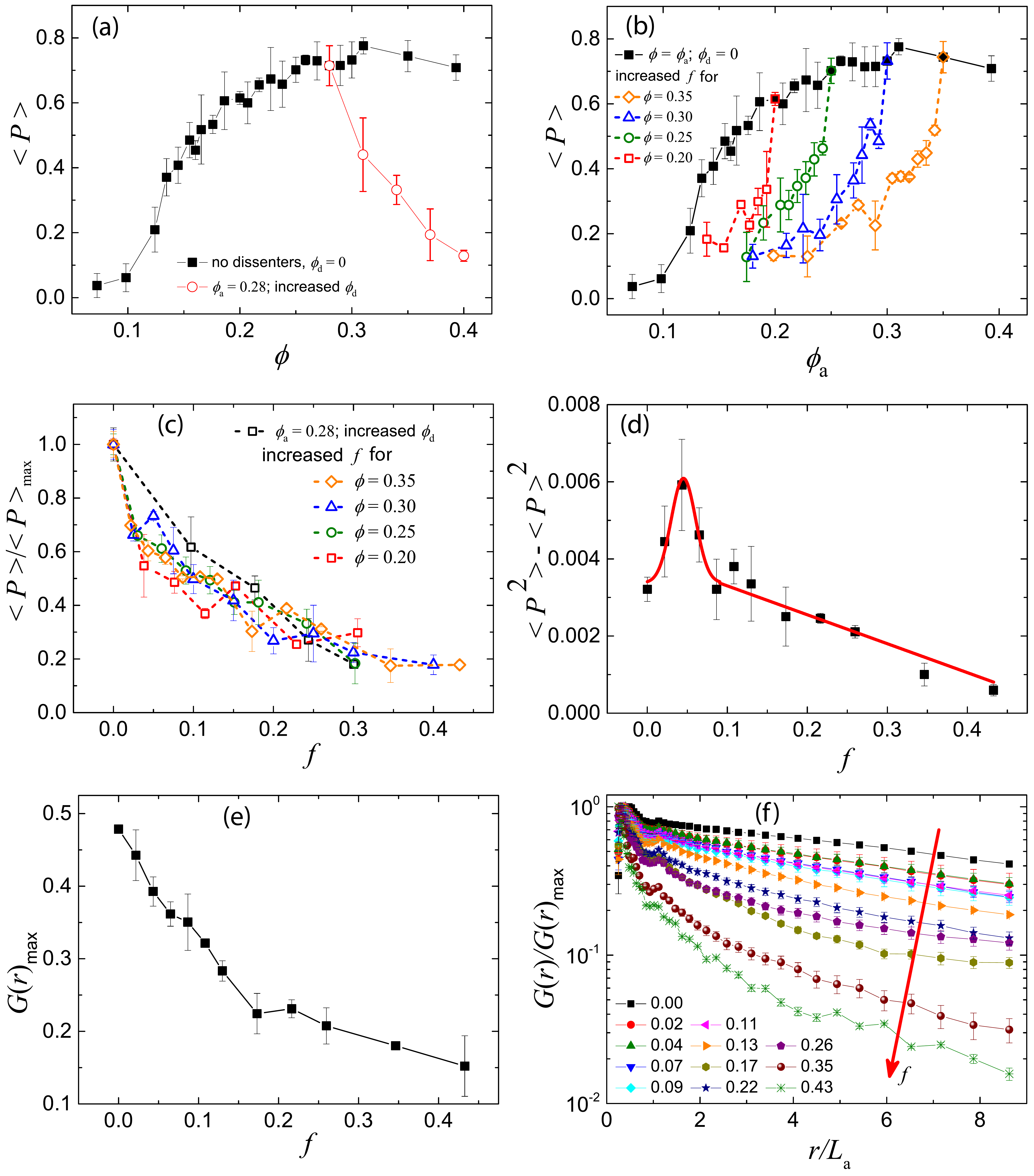}
\caption{The system with aligners + dissenters; error bars are calculated in the same way as in Fig.~\ref{F2}. (a) $\left\langle P \right\rangle$ vs $\phi$ is shown along with the plot from Fig.~\ref{F2}(b). $\phi=\phi_\mathrm{a}+\phi_\mathrm{d}$ where $\phi_\mathrm{a}$ is the aligners' area fraction and $\phi_\mathrm{d}$ is the dissenters' area fraction. (b) For four different fixed $\phi$, $\left\langle P \right\rangle$ vs $\phi_\mathrm{a}$ plots along with the plot from Fig.~\ref{F2}(b). The dissenter fraction $f=\phi_\mathrm{d}/\phi$ was increased until the disordered state was reached. (c) $\left\langle P \right\rangle$ normalized w.r.t. the maxima are plotted vs $f$ for all the mixed systems. For $\phi=0.35$, (d) the variance of $P$ in the steady-state is plotted vs $f$. The red curve is the guide to the eyes, indicating a maximum near $f = 0.05$. For $\phi=0.35$, (e) the observed $G(r)$ maxima vs $f$ and (f) $G(r)/G(r)_\mathrm{max}$ vs $r/L_\mathrm{a}$ for different $f$ are shown, without including the dissenters in the calculation. The red arrow is towards increasing $f$.}
\label{F3}
\end{figure*}

We next discuss the effects of dissenters on the flocking. Figure~\ref{F3} broadly summarizes our experimental findings with the aligner-dissenter mixed system. In the mixed system, the area fraction of aligners and the area fraction of dissenters are denoted by $\phi_\mathrm{a}$ and $\phi_\mathrm{d}$ respectively, ($\phi=\phi_\mathrm{a}+\phi_\mathrm{d}$). The trapping and sorting study with these active particles (at low $\phi$) and having a trap in the cell is reported elsewhere \cite{kumar2019trapping}. At first, to see the effect of dissenters on the flocking behavior of aligners, dissenters are added with an increasing number to increase total $\phi$, keeping the aligners area fraction constant ($\phi_\mathrm{a}=0.28$), and the mixture is dispersed in the cell to follow the dynamics in time. The presence of dissenters disrupts the flocking (Fig.~\ref{F3}(a)), and the system shows reduced order parameter $\left\langle P \right\rangle$ with increasing $\phi$ (as we increase $\phi_\mathrm{d}$) (see Supplemental Material \cite{supmat}, Movie S8). All error bars in Fig.~\ref{F3} are the standard deviations over three repeated experiments. At a higher $\phi_\mathrm{d}$, it shows completely disordered motion (see Supplemental Material \cite{supmat}, Movie S9). We note that the mixed system phase segregates at the high value of $\phi$ ($\geq$ 0.58) and shows the active jammed state for $\phi \geq 0.70$. 

Now we keep the total $\phi$ fixed and increase the dissenters' fraction in the system ($f=\phi_\mathrm{d}/\phi$) by substituting some aligners with dissenters in the system (see Supplemental Material \cite{supmat}, Movies S10, S11). In Fig.~\ref{F3}(b), we plot $\left\langle P \right\rangle$ vs $\phi_\mathrm{a}$ with increasing $f$ for different fixed values of $\phi$ until the system shows the completely disordered state. Black filled squares represent the same data as in Fig.~\ref{F2}(b), where we have $\phi_\mathrm{a}=\phi$ as $\phi_\mathrm{d}=0$. Compared to the system with pure aligners, the cross over from ordered to disordered state happens at higher values of $\phi_\mathrm{a}$ (depending on starting $\phi$), implying that the effect of the dissenters is much stronger than that of simply diluting the system. To quantify these effects in terms of the fraction of dissenter particles $f$, we plot normalized $\left\langle P \right\rangle$ vs $f$ in Fig.~\ref{F3}(c) which shows the collapse of all the data sets. We expect that the data collapse may be better with much larger system size. The normalized order parameter $\left\langle P \right\rangle/\left\langle P \right\rangle_\mathrm{max}$ decreases continuously from 1 and reaches the completely disordered state nearly at $f \sim 0.3$. Figure~\ref{F3}(c) suggests that the relative suppression of flocking by dissenters is independent of the area fraction of aligners. 

The variance of the steady-state order parameter would be proportional to the susceptibility if this was an equilibrium system, and is in any case a measure of the magnitude of fluctuations. In the simulation study on the mixed aligner-dissenter system \cite{yllanes2017many}, Yllanes et al have shown that the variance shows a maximum at a dissenters' fraction where the order parameter reduces to $\sim 0.5$ of its maximum value, indicative of an underlying phase transition in the infinite-size limit. We were curious to see if the variance of $P$ exhibits such non-monotonic behavior with respect to $f$. Fig.~\ref{F3}(d) plots the variance of $P$ as a function of $f$ calculated considering the fluctuations observed in the steady-state for $\phi=0.35$, showing a peak at $f \sim 0.05$ where the order parameter is close to 0.5$\left\langle P \right\rangle_\mathrm{max}$, similar to the simulation results (see Fig. 3 of Ref. \cite{yllanes2017many}). Here we may add a word of caution that the crossover point extracted from our data is only indicative due to finite size effects.

It is also interesting to look at the mixed systems without including the contributions of dissenters in calculating the $\left\langle P \right\rangle$, $G(r)$, etc. We have estimated separately contributions of aligners and dissenters in the normalized $\left\langle P \right\rangle$ vs $f$ for $\phi=0.35$ and observed negligible contribution of dissenters (see Supplemental Material \cite{supmat}, Fig.~S5(a)). Also, the variance of $P$ vs $f$ considering only aligners (Supplemental Material \cite{supmat}, Fig.~S5(b)) shows reduced values but still retains the maxima close to $f=0.05$ as in Fig.~\ref{F3}(d). The negligible contribution of dissenters in the normalized $\left\langle P \right\rangle$ and variance is expected as they do not contribute to the collective behavior.

\begin{figure}
\includegraphics[width=0.5\textwidth]{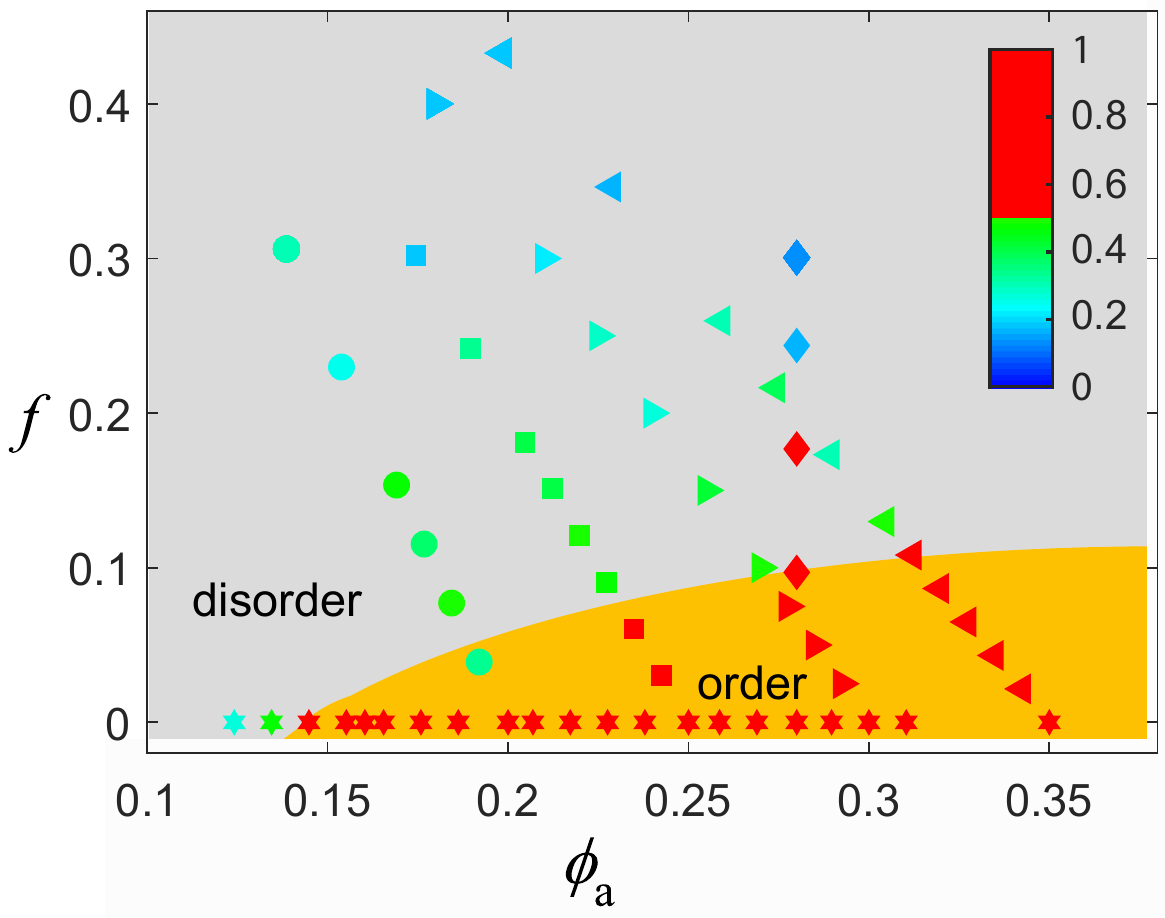}
\caption{The flocking phase diagram in the $f \mbox{-} \phi_\mathrm{a}$ plane, considering data sets of Fig.~\ref{F2} and Fig.~\ref{F3}, represented by different symbols. Color bars represent $\left\langle P \right\rangle/\left\langle P \right\rangle_\mathrm{max}$ value. The order region and the disorder region are indicated by considering the crossover value $\left\langle P \right\rangle/\left\langle P \right\rangle_\mathrm{max} \sim 0.5$.}
\label{F4}
\end{figure}

Next, we quantify the particle-level interactions in the mixed systems. For $\phi = 0.35$, we plot $G(r)_\mathrm{max}$ vs $f$ and $G(r)/G(r)_\mathrm{max}$ vs $r/L_\mathrm{a}$ in Fig.~\ref{F3}(e), (f) respectively for only aligners contribution without taking dissenters (also see Supplemental Material \cite{supmat}, Fig.~S6). The plot shows a monotonic evolution from highly correlated to low correlated state. When a small fraction of dissenters is present, the system can retain some order. With high dissenters fraction, $G(r)/G(r)_\mathrm{max}$ falls rapidly with $r/L_\mathrm{a}$ to a low value showing negligible correlation between aligners. The dissenter-dissenter pairs always show low orientational correlation even when the system shows flocking with high $\left\langle P \right\rangle$ at low $f$. At high $f$, aligner-aligner pairs eventually show low orientational correlation due to the disturbance from the dissenters (see Supplemental Material \cite{supmat}, Fig.~S7). By combining the observations, shown in Fig.~\ref{F2}(b) and Fig.~\ref{F3}(a),(b), the phase diagram thus constructed of this flocking ordered to isotropic disordered transition is shown in Fig.~\ref{F4}. Here we take $\left\langle P \right\rangle/\left\langle P \right\rangle_\mathrm{max} \sim 0.5$ to mark the boundary between order and disorder regions.

\section{CONCLUSIONS}
In summary, our dry active granular system with two-step-tapered polar rods shows flocking in a wide range of the area fraction without any intervening medium. The order parameter vary smoothly around $\phi_c$ due to finite system size which would be discontinuous in the larger system sizes as shown in the simulations \cite{chate2019dry}. Orientational correlations grow as the system approaches a threshold area fraction. We experimentally realize motile ``dissenters'' in the form of one-step-tapered polar particles that move in a much more noisy fashion than the two-step-tapered particles. To follow the transient behavior of the system during flocked to de-flocking transition in presence of dissenters, one has to pause the shaker after flock formation and replace some aligners in random places by dissenters keeping all positions and directions unchanged, then run the shaker to follow the dynamics in time which is a very laborious task. Rather, we mix and disperse them to follow the system in the steady-state. The effect of the dissenters is much stronger than that of simply diluting the system and it depends only on the ordering in the system, not on the aligners area fractions. We have shown that the dissenters disrupt the flocking of the active granular matter by introducing more noise to the system and destroying the orientational correlation between aligners. The peak in the variance of the system's order parameter at $f \sim 0.05$ is associated with the finite size crossover from order to disordered state. Our experimental results can be visualized in terms of real-life examples. For example, flocking is observed in animal groups having a small fraction of baby animals along with the adults, whereas the flocking is absent with the large fraction of the baby animals. Another situation where de-flocking can be desirable is the motion of a crowd in high-risk situations where a large number of dissenters do not allow collective motion. It will be interesting to study the role of apolar rods as dissenters in flocking of the aligners.

\section{ACKNOWLEDGMENTS}
We thank Prof. Sriram Ramaswamy for critical reading of the manuscript. A.K.S. thanks Department of Science and Technology (DST), India for the support through Year of Science Professorship. P.K.B. thanks University Grants Commission (UGC) for the Senior Research Fellowship.

\bibliography{refsmotiledissenters}

\begin{thebibliography}{27}%
\makeatletter
\providecommand \@ifxundefined [1]{%
 \@ifx{#1\undefined}
}%
\providecommand \@ifnum [1]{%
 \ifnum #1\expandafter \@firstoftwo
 \else \expandafter \@secondoftwo
 \fi
}%
\providecommand \@ifx [1]{%
 \ifx #1\expandafter \@firstoftwo
 \else \expandafter \@secondoftwo
 \fi
}%
\providecommand \natexlab [1]{#1}%
\providecommand \enquote  [1]{``#1''}%
\providecommand \bibnamefont  [1]{#1}%
\providecommand \bibfnamefont [1]{#1}%
\providecommand \citenamefont [1]{#1}%
\providecommand \href@noop [0]{\@secondoftwo}%
\providecommand \href [0]{\begingroup \@sanitize@url \@href}%
\providecommand \@href[1]{\@@startlink{#1}\@@href}%
\providecommand \@@href[1]{\endgroup#1\@@endlink}%
\providecommand \@sanitize@url [0]{\catcode `\\12\catcode `\$12\catcode
  `\&12\catcode `\#12\catcode `\^12\catcode `\_12\catcode `\%12\relax}%
\providecommand \@@startlink[1]{}%
\providecommand \@@endlink[0]{}%
\providecommand \url  [0]{\begingroup\@sanitize@url \@url }%
\providecommand \@url [1]{\endgroup\@href {#1}{\urlprefix }}%
\providecommand \urlprefix  [0]{URL }%
\providecommand \Eprint [0]{\href }%
\providecommand \doibase [0]{http://dx.doi.org/}%
\providecommand \selectlanguage [0]{\@gobble}%
\providecommand \bibinfo  [0]{\@secondoftwo}%
\providecommand \bibfield  [0]{\@secondoftwo}%
\providecommand \translation [1]{[#1]}%
\providecommand \BibitemOpen [0]{}%
\providecommand \bibitemStop [0]{}%
\providecommand \bibitemNoStop [0]{.\EOS\space}%
\providecommand \EOS [0]{\spacefactor3000\relax}%
\providecommand \BibitemShut  [1]{\csname bibitem#1\endcsname}%
\let\auto@bib@innerbib\@empty
\bibitem [{\citenamefont {Yllanes}\ \emph {et~al.}(2017)\citenamefont
  {Yllanes}, \citenamefont {Leoni},\ and\ \citenamefont
  {Marchetti}}]{yllanes2017many}%
  \BibitemOpen
  \bibfield  {author} {\bibinfo {author} {\bibfnamefont {D.}~\bibnamefont
  {Yllanes}}, \bibinfo {author} {\bibfnamefont {M.}~\bibnamefont {Leoni}}, \
  and\ \bibinfo {author} {\bibfnamefont {M.~C.}\ \bibnamefont {Marchetti}},\
  }\href@noop {} {\bibfield  {journal} {\bibinfo  {journal} {New J. Phys.}\
  }\textbf {\bibinfo {volume} {19}},\ \bibinfo {pages} {103026} (\bibinfo
  {year} {2017})}\BibitemShut {NoStop}%
\bibitem [{\citenamefont {Chepizhko}\ \emph {et~al.}(2013)\citenamefont
  {Chepizhko}, \citenamefont {Altmann},\ and\ \citenamefont
  {Peruani}}]{chepizhko2013optimal}%
  \BibitemOpen
  \bibfield  {author} {\bibinfo {author} {\bibfnamefont {O.}~\bibnamefont
  {Chepizhko}}, \bibinfo {author} {\bibfnamefont {E.~G.}\ \bibnamefont
  {Altmann}}, \ and\ \bibinfo {author} {\bibfnamefont {F.}~\bibnamefont
  {Peruani}},\ }\href@noop {} {\bibfield  {journal} {\bibinfo  {journal} {Phys.
  Rev. Lett.}\ }\textbf {\bibinfo {volume} {110}},\ \bibinfo {pages} {238101}
  (\bibinfo {year} {2013})}\BibitemShut {NoStop}%
\bibitem [{\citenamefont {Berdahl}\ \emph {et~al.}(2013)\citenamefont
  {Berdahl}, \citenamefont {Torney}, \citenamefont {Ioannou}, \citenamefont
  {Faria},\ and\ \citenamefont {Couzin}}]{berdahl2013emergent}%
  \BibitemOpen
  \bibfield  {author} {\bibinfo {author} {\bibfnamefont {A.}~\bibnamefont
  {Berdahl}}, \bibinfo {author} {\bibfnamefont {C.~J.}\ \bibnamefont {Torney}},
  \bibinfo {author} {\bibfnamefont {C.~C.}\ \bibnamefont {Ioannou}}, \bibinfo
  {author} {\bibfnamefont {J.~J.}\ \bibnamefont {Faria}}, \ and\ \bibinfo
  {author} {\bibfnamefont {I.~D.}\ \bibnamefont {Couzin}},\ }\href@noop {}
  {\bibfield  {journal} {\bibinfo  {journal} {Science}\ }\textbf {\bibinfo
  {volume} {339}},\ \bibinfo {pages} {574} (\bibinfo {year}
  {2013})}\BibitemShut {NoStop}%
\bibitem [{\citenamefont {Quint}\ and\ \citenamefont
  {Gopinathan}(2015)}]{quint2015topologically}%
  \BibitemOpen
  \bibfield  {author} {\bibinfo {author} {\bibfnamefont {D.~A.}\ \bibnamefont
  {Quint}}\ and\ \bibinfo {author} {\bibfnamefont {A.}~\bibnamefont
  {Gopinathan}},\ }\href@noop {} {\bibfield  {journal} {\bibinfo  {journal}
  {Phys. Biol.}\ }\textbf {\bibinfo {volume} {12}},\ \bibinfo {pages} {046008}
  (\bibinfo {year} {2015})}\BibitemShut {NoStop}%
\bibitem [{\citenamefont {Pin{\c{c}}e}\ \emph {et~al.}(2016)\citenamefont
  {Pin{\c{c}}e}, \citenamefont {Velu}, \citenamefont {Callegari}, \citenamefont
  {Elahi}, \citenamefont {Gigan}, \citenamefont {Volpe},\ and\ \citenamefont
  {Volpe}}]{pincce2016disorder}%
  \BibitemOpen
  \bibfield  {author} {\bibinfo {author} {\bibfnamefont {E.}~\bibnamefont
  {Pin{\c{c}}e}}, \bibinfo {author} {\bibfnamefont {S.~K.~P.}\ \bibnamefont
  {Velu}}, \bibinfo {author} {\bibfnamefont {A.}~\bibnamefont {Callegari}},
  \bibinfo {author} {\bibfnamefont {P.}~\bibnamefont {Elahi}}, \bibinfo
  {author} {\bibfnamefont {S.}~\bibnamefont {Gigan}}, \bibinfo {author}
  {\bibfnamefont {G.}~\bibnamefont {Volpe}}, \ and\ \bibinfo {author}
  {\bibfnamefont {G.}~\bibnamefont {Volpe}},\ }\href@noop {} {\bibfield
  {journal} {\bibinfo  {journal} {Nat. Commun.}\ }\textbf {\bibinfo {volume}
  {7}},\ \bibinfo {pages} {10907} (\bibinfo {year} {2016})}\BibitemShut
  {NoStop}%
\bibitem [{\citenamefont {Morin}\ \emph {et~al.}(2017)\citenamefont {Morin},
  \citenamefont {Desreumaux}, \citenamefont {Caussin},\ and\ \citenamefont
  {Bartolo}}]{morin2017distortion}%
  \BibitemOpen
  \bibfield  {author} {\bibinfo {author} {\bibfnamefont {A.}~\bibnamefont
  {Morin}}, \bibinfo {author} {\bibfnamefont {N.}~\bibnamefont {Desreumaux}},
  \bibinfo {author} {\bibfnamefont {J.-B.}\ \bibnamefont {Caussin}}, \ and\
  \bibinfo {author} {\bibfnamefont {D.}~\bibnamefont {Bartolo}},\ }\href@noop
  {} {\bibfield  {journal} {\bibinfo  {journal} {Nat. Phys.}\ }\textbf
  {\bibinfo {volume} {13}},\ \bibinfo {pages} {63} (\bibinfo {year}
  {2017})}\BibitemShut {NoStop}%
\bibitem [{\citenamefont {S{\'a}ndor}\ \emph {et~al.}(2017)\citenamefont
  {S{\'a}ndor}, \citenamefont {Libal}, \citenamefont {Reichhardt},\ and\
  \citenamefont {Reichhardt}}]{sandor2017dynamic}%
  \BibitemOpen
  \bibfield  {author} {\bibinfo {author} {\bibfnamefont {C.}~\bibnamefont
  {S{\'a}ndor}}, \bibinfo {author} {\bibfnamefont {A.}~\bibnamefont {Libal}},
  \bibinfo {author} {\bibfnamefont {C.}~\bibnamefont {Reichhardt}}, \ and\
  \bibinfo {author} {\bibfnamefont {C.~J.~O.}\ \bibnamefont {Reichhardt}},\
  }\href@noop {} {\bibfield  {journal} {\bibinfo  {journal} {Phys. Rev. E}\
  }\textbf {\bibinfo {volume} {95}},\ \bibinfo {pages} {032606} (\bibinfo
  {year} {2017})}\BibitemShut {NoStop}%
\bibitem [{\citenamefont {Reichhardt}\ and\ \citenamefont
  {Reichhardt}(2017)}]{reichhardt2017collective}%
  \BibitemOpen
  \bibfield  {author} {\bibinfo {author} {\bibfnamefont {C.~J.~O.}\
  \bibnamefont {Reichhardt}}\ and\ \bibinfo {author} {\bibfnamefont
  {C.}~\bibnamefont {Reichhardt}},\ }\href@noop {} {\bibfield  {journal}
  {\bibinfo  {journal} {Nat. Phys.}\ }\textbf {\bibinfo {volume} {13}},\
  \bibinfo {pages} {10} (\bibinfo {year} {2017})}\BibitemShut {NoStop}%
\bibitem [{\citenamefont {Reichhardt}\ and\ \citenamefont
  {Reichhardt}(2018)}]{reichhardt2018clogging}%
  \BibitemOpen
  \bibfield  {author} {\bibinfo {author} {\bibfnamefont {C.}~\bibnamefont
  {Reichhardt}}\ and\ \bibinfo {author} {\bibfnamefont {C.~J.~O.}\ \bibnamefont
  {Reichhardt}},\ }\href@noop {} {\bibfield  {journal} {\bibinfo  {journal}
  {Phys. Rev. E}\ }\textbf {\bibinfo {volume} {97}},\ \bibinfo {pages} {052613}
  (\bibinfo {year} {2018})}\BibitemShut {NoStop}%
\bibitem [{\citenamefont {Das}\ \emph {et~al.}(2018)\citenamefont {Das},
  \citenamefont {Kumar},\ and\ \citenamefont {Mishra}}]{das2018polar}%
  \BibitemOpen
  \bibfield  {author} {\bibinfo {author} {\bibfnamefont {R.}~\bibnamefont
  {Das}}, \bibinfo {author} {\bibfnamefont {M.}~\bibnamefont {Kumar}}, \ and\
  \bibinfo {author} {\bibfnamefont {S.}~\bibnamefont {Mishra}},\ }\href@noop {}
  {\bibfield  {journal} {\bibinfo  {journal} {Phys. Rev. E}\ }\textbf {\bibinfo
  {volume} {98}},\ \bibinfo {pages} {060602} (\bibinfo {year}
  {2018})}\BibitemShut {NoStop}%
\bibitem [{\citenamefont {Ai}\ \emph {et~al.}(2019)\citenamefont {Ai},
  \citenamefont {Meng}, \citenamefont {He},\ and\ \citenamefont
  {Zhang}}]{ai2019flow}%
  \BibitemOpen
  \bibfield  {author} {\bibinfo {author} {\bibfnamefont {B.-q.}\ \bibnamefont
  {Ai}}, \bibinfo {author} {\bibfnamefont {F.-h.}\ \bibnamefont {Meng}},
  \bibinfo {author} {\bibfnamefont {Y.-l.}\ \bibnamefont {He}}, \ and\ \bibinfo
  {author} {\bibfnamefont {X.-m.}\ \bibnamefont {Zhang}},\ }\href@noop {}
  {\bibfield  {journal} {\bibinfo  {journal} {Soft Matter}\ }\textbf {\bibinfo
  {volume} {15}},\ \bibinfo {pages} {3443} (\bibinfo {year}
  {2019})}\BibitemShut {NoStop}%
\bibitem [{\citenamefont {Guttal}\ and\ \citenamefont
  {Couzin}(2010)}]{guttal2010social}%
  \BibitemOpen
  \bibfield  {author} {\bibinfo {author} {\bibfnamefont {V.}~\bibnamefont
  {Guttal}}\ and\ \bibinfo {author} {\bibfnamefont {I.~D.}\ \bibnamefont
  {Couzin}},\ }\href@noop {} {\bibfield  {journal} {\bibinfo  {journal} {Proc.
  Natl. Acad. Sci. U.S.A.}\ }\textbf {\bibinfo {volume} {107}},\ \bibinfo
  {pages} {16172} (\bibinfo {year} {2010})}\BibitemShut {NoStop}%
\bibitem [{\citenamefont {Couzin}\ \emph {et~al.}(2011)\citenamefont {Couzin},
  \citenamefont {Ioannou}, \citenamefont {Demirel}, \citenamefont {Gross},
  \citenamefont {Torney}, \citenamefont {Hartnett}, \citenamefont {Conradt},
  \citenamefont {Levin},\ and\ \citenamefont {Leonard}}]{couzin2011uninformed}%
  \BibitemOpen
  \bibfield  {author} {\bibinfo {author} {\bibfnamefont {I.~D.}\ \bibnamefont
  {Couzin}}, \bibinfo {author} {\bibfnamefont {C.~C.}\ \bibnamefont {Ioannou}},
  \bibinfo {author} {\bibfnamefont {G.}~\bibnamefont {Demirel}}, \bibinfo
  {author} {\bibfnamefont {T.}~\bibnamefont {Gross}}, \bibinfo {author}
  {\bibfnamefont {C.~J.}\ \bibnamefont {Torney}}, \bibinfo {author}
  {\bibfnamefont {A.}~\bibnamefont {Hartnett}}, \bibinfo {author}
  {\bibfnamefont {L.}~\bibnamefont {Conradt}}, \bibinfo {author} {\bibfnamefont
  {S.~A.}\ \bibnamefont {Levin}}, \ and\ \bibinfo {author} {\bibfnamefont
  {N.~E.}\ \bibnamefont {Leonard}},\ }\href@noop {} {\bibfield  {journal}
  {\bibinfo  {journal} {Science}\ }\textbf {\bibinfo {volume} {334}},\ \bibinfo
  {pages} {1578} (\bibinfo {year} {2011})}\BibitemShut {NoStop}%
\bibitem [{\citenamefont {McCandlish}\ \emph {et~al.}(2012)\citenamefont
  {McCandlish}, \citenamefont {Baskaran},\ and\ \citenamefont
  {Hagan}}]{mccandlish2012spontaneous}%
  \BibitemOpen
  \bibfield  {author} {\bibinfo {author} {\bibfnamefont {S.~R.}\ \bibnamefont
  {McCandlish}}, \bibinfo {author} {\bibfnamefont {A.}~\bibnamefont
  {Baskaran}}, \ and\ \bibinfo {author} {\bibfnamefont {M.~F.}\ \bibnamefont
  {Hagan}},\ }\href@noop {} {\bibfield  {journal} {\bibinfo  {journal} {Soft
  Matter}\ }\textbf {\bibinfo {volume} {8}},\ \bibinfo {pages} {2527} (\bibinfo
  {year} {2012})}\BibitemShut {NoStop}%
\bibitem [{\citenamefont {Baglietto}\ \emph {et~al.}(2013)\citenamefont
  {Baglietto}, \citenamefont {Albano},\ and\ \citenamefont
  {Candia}}]{baglietto2013gregarious}%
  \BibitemOpen
  \bibfield  {author} {\bibinfo {author} {\bibfnamefont {G.}~\bibnamefont
  {Baglietto}}, \bibinfo {author} {\bibfnamefont {E.~V.}\ \bibnamefont
  {Albano}}, \ and\ \bibinfo {author} {\bibfnamefont {J.}~\bibnamefont
  {Candia}},\ }\href@noop {} {\bibfield  {journal} {\bibinfo  {journal}
  {Physica A}\ }\textbf {\bibinfo {volume} {392}},\ \bibinfo {pages} {3240}
  (\bibinfo {year} {2013})}\BibitemShut {NoStop}%
\bibitem [{\citenamefont {Ariel}\ \emph {et~al.}(2015)\citenamefont {Ariel},
  \citenamefont {Rimer},\ and\ \citenamefont {Ben-Jacob}}]{ariel2015order}%
  \BibitemOpen
  \bibfield  {author} {\bibinfo {author} {\bibfnamefont {G.}~\bibnamefont
  {Ariel}}, \bibinfo {author} {\bibfnamefont {O.}~\bibnamefont {Rimer}}, \ and\
  \bibinfo {author} {\bibfnamefont {E.}~\bibnamefont {Ben-Jacob}},\ }\href@noop
  {} {\bibfield  {journal} {\bibinfo  {journal} {J. Stat. Phys.}\ }\textbf
  {\bibinfo {volume} {158}},\ \bibinfo {pages} {579} (\bibinfo {year}
  {2015})}\BibitemShut {NoStop}%
\bibitem [{\citenamefont {Copenhagen}\ \emph {et~al.}(2016)\citenamefont
  {Copenhagen}, \citenamefont {Quint},\ and\ \citenamefont
  {Gopinathan}}]{copenhagen2016self}%
  \BibitemOpen
  \bibfield  {author} {\bibinfo {author} {\bibfnamefont {K.}~\bibnamefont
  {Copenhagen}}, \bibinfo {author} {\bibfnamefont {D.~A.}\ \bibnamefont
  {Quint}}, \ and\ \bibinfo {author} {\bibfnamefont {A.}~\bibnamefont
  {Gopinathan}},\ }\href@noop {} {\bibfield  {journal} {\bibinfo  {journal}
  {Sci. Rep.}\ }\textbf {\bibinfo {volume} {6}},\ \bibinfo {pages} {31808}
  (\bibinfo {year} {2016})}\BibitemShut {NoStop}%
\bibitem [{sup()}]{supmat}%
  \BibitemOpen
  \href@noop {} {\enquote {\bibinfo {title} {See supplemental material at this
  url, which includes; legends to movies, setup calibration, image analysis,
  refs. and supplemental figures.}}\ }\bibinfo {howpublished}
  {\url{https://drive.google.com/open?id=1n-egjVPLSZ-CojMPCFWKRrNkaxBuhmHb}}\BibitemShut
  {NoStop}%
\bibitem [{\citenamefont {Yamada}\ \emph {et~al.}(2003)\citenamefont {Yamada},
  \citenamefont {Hondou},\ and\ \citenamefont {Sano}}]{yamada2003coherent}%
  \BibitemOpen
  \bibfield  {author} {\bibinfo {author} {\bibfnamefont {D.}~\bibnamefont
  {Yamada}}, \bibinfo {author} {\bibfnamefont {T.}~\bibnamefont {Hondou}}, \
  and\ \bibinfo {author} {\bibfnamefont {M.}~\bibnamefont {Sano}},\ }\href@noop
  {} {\bibfield  {journal} {\bibinfo  {journal} {Phys. Rev. E}\ }\textbf
  {\bibinfo {volume} {67}},\ \bibinfo {pages} {040301} (\bibinfo {year}
  {2003})}\BibitemShut {NoStop}%
\bibitem [{\citenamefont {Narayan}\ \emph {et~al.}(2007)\citenamefont
  {Narayan}, \citenamefont {Ramaswamy},\ and\ \citenamefont
  {Menon}}]{narayan2007long}%
  \BibitemOpen
  \bibfield  {author} {\bibinfo {author} {\bibfnamefont {V.}~\bibnamefont
  {Narayan}}, \bibinfo {author} {\bibfnamefont {S.}~\bibnamefont {Ramaswamy}},
  \ and\ \bibinfo {author} {\bibfnamefont {N.}~\bibnamefont {Menon}},\
  }\href@noop {} {\bibfield  {journal} {\bibinfo  {journal} {Science}\ }\textbf
  {\bibinfo {volume} {317}},\ \bibinfo {pages} {105} (\bibinfo {year}
  {2007})}\BibitemShut {NoStop}%
\bibitem [{\citenamefont {Kumar}\ \emph {et~al.}(2011)\citenamefont {Kumar},
  \citenamefont {Ramaswamy},\ and\ \citenamefont {Sood}}]{kumar2011symmetry}%
  \BibitemOpen
  \bibfield  {author} {\bibinfo {author} {\bibfnamefont {N.}~\bibnamefont
  {Kumar}}, \bibinfo {author} {\bibfnamefont {S.}~\bibnamefont {Ramaswamy}}, \
  and\ \bibinfo {author} {\bibfnamefont {A.~K.}\ \bibnamefont {Sood}},\
  }\href@noop {} {\bibfield  {journal} {\bibinfo  {journal} {Phys. Rev. Lett.}\
  }\textbf {\bibinfo {volume} {106}},\ \bibinfo {pages} {118001} (\bibinfo
  {year} {2011})}\BibitemShut {NoStop}%
\bibitem [{\citenamefont {Kumar}\ \emph {et~al.}(2019)\citenamefont {Kumar},
  \citenamefont {Gupta}, \citenamefont {Soni}, \citenamefont {Ramaswamy},\ and\
  \citenamefont {Sood}}]{kumar2019trapping}%
  \BibitemOpen
  \bibfield  {author} {\bibinfo {author} {\bibfnamefont {N.}~\bibnamefont
  {Kumar}}, \bibinfo {author} {\bibfnamefont {R.~K.}\ \bibnamefont {Gupta}},
  \bibinfo {author} {\bibfnamefont {H.}~\bibnamefont {Soni}}, \bibinfo {author}
  {\bibfnamefont {S.}~\bibnamefont {Ramaswamy}}, \ and\ \bibinfo {author}
  {\bibfnamefont {A.~K.}\ \bibnamefont {Sood}},\ }\href@noop {} {\bibfield
  {journal} {\bibinfo  {journal} {Phys. Rev. E}\ }\textbf {\bibinfo {volume}
  {99}},\ \bibinfo {pages} {032605} (\bibinfo {year} {2019})}\BibitemShut
  {NoStop}%
\bibitem [{\citenamefont {Schindelin}\ \emph {et~al.}(2012)\citenamefont
  {Schindelin}, \citenamefont {Arganda-Carreras}, \citenamefont {Frise},
  \citenamefont {Kaynig}, \citenamefont {Longair}, \citenamefont {Pietzsch},
  \citenamefont {Preibisch}, \citenamefont {Rueden}, \citenamefont {Saalfeld},\
  and\ \citenamefont {et~al.}}]{schindelin2012fiji}%
  \BibitemOpen
  \bibfield  {author} {\bibinfo {author} {\bibfnamefont {J.}~\bibnamefont
  {Schindelin}}, \bibinfo {author} {\bibfnamefont {I.}~\bibnamefont
  {Arganda-Carreras}}, \bibinfo {author} {\bibfnamefont {E.}~\bibnamefont
  {Frise}}, \bibinfo {author} {\bibfnamefont {V.}~\bibnamefont {Kaynig}},
  \bibinfo {author} {\bibfnamefont {M.}~\bibnamefont {Longair}}, \bibinfo
  {author} {\bibfnamefont {T.}~\bibnamefont {Pietzsch}}, \bibinfo {author}
  {\bibfnamefont {S.}~\bibnamefont {Preibisch}}, \bibinfo {author}
  {\bibfnamefont {C.}~\bibnamefont {Rueden}}, \bibinfo {author} {\bibfnamefont
  {S.}~\bibnamefont {Saalfeld}}, \ and\ \bibinfo {author} {\bibnamefont
  {et~al.}},\ }\href@noop {} {\bibfield  {journal} {\bibinfo  {journal} {Nat.
  Methods}\ }\textbf {\bibinfo {volume} {9}},\ \bibinfo {pages} {676} (\bibinfo
  {year} {2012})}\BibitemShut {NoStop}%
\bibitem [{\citenamefont {Henkes}\ \emph {et~al.}(2011)\citenamefont {Henkes},
  \citenamefont {Fily},\ and\ \citenamefont {Marchetti}}]{henkes2011active}%
  \BibitemOpen
  \bibfield  {author} {\bibinfo {author} {\bibfnamefont {S.}~\bibnamefont
  {Henkes}}, \bibinfo {author} {\bibfnamefont {Y.}~\bibnamefont {Fily}}, \ and\
  \bibinfo {author} {\bibfnamefont {M.~C.}\ \bibnamefont {Marchetti}},\
  }\href@noop {} {\bibfield  {journal} {\bibinfo  {journal} {Phys. Rev. E}\
  }\textbf {\bibinfo {volume} {84}},\ \bibinfo {pages} {040301} (\bibinfo
  {year} {2011})}\BibitemShut {NoStop}%
\bibitem [{\citenamefont {Kumar}\ \emph {et~al.}(2014)\citenamefont {Kumar},
  \citenamefont {Soni}, \citenamefont {Ramaswamy},\ and\ \citenamefont
  {Sood}}]{kumar2014flocking}%
  \BibitemOpen
  \bibfield  {author} {\bibinfo {author} {\bibfnamefont {N.}~\bibnamefont
  {Kumar}}, \bibinfo {author} {\bibfnamefont {H.}~\bibnamefont {Soni}},
  \bibinfo {author} {\bibfnamefont {S.}~\bibnamefont {Ramaswamy}}, \ and\
  \bibinfo {author} {\bibfnamefont {A.~K.}\ \bibnamefont {Sood}},\ }\href@noop
  {} {\bibfield  {journal} {\bibinfo  {journal} {Nat. Commun.}\ }\textbf
  {\bibinfo {volume} {5}},\ \bibinfo {pages} {4688} (\bibinfo {year}
  {2014})}\BibitemShut {NoStop}%
\bibitem [{\citenamefont {Chat{\'e}}\ and\ \citenamefont
  {Mahault}(2019)}]{chate2019dry}%
  \BibitemOpen
  \bibfield  {author} {\bibinfo {author} {\bibfnamefont {H.}~\bibnamefont
  {Chat{\'e}}}\ and\ \bibinfo {author} {\bibfnamefont {B.}~\bibnamefont
  {Mahault}},\ }\href@noop {} {\bibfield  {journal} {\bibinfo  {journal}
  {\textit{Dry, aligning, dilute, active matter: A synthetic and self-contained
  overview}, arXiv:1906.05542}\ } (\bibinfo {year} {2019})}\BibitemShut
  {NoStop}%
\bibitem [{\citenamefont {Reichhardt}\ and\ \citenamefont
  {Reichhardt}(2014)}]{reichhardt2014absorbing}%
  \BibitemOpen
  \bibfield  {author} {\bibinfo {author} {\bibfnamefont {C.}~\bibnamefont
  {Reichhardt}}\ and\ \bibinfo {author} {\bibfnamefont {C.~J.~O.}\ \bibnamefont
  {Reichhardt}},\ }\href@noop {} {\bibfield  {journal} {\bibinfo  {journal}
  {Soft Matter}\ }\textbf {\bibinfo {volume} {10}},\ \bibinfo {pages} {7502}
  (\bibinfo {year} {2014})}\BibitemShut {NoStop}%
\end{thebibliography}%

\newpage

\newpage

\newpage

\newpage

\end{document}